\documentclass[]{iopart}
\usepackage{graphicx}
\begin{document}
\title{Antiferro and ferromagnetic ordering in PrGe single crystal}
\author{Pranab Kumar Das, K. Ramesh Kumar, R. Kulkarni, S. K. Dhar and A. Thamizhavel}
\address{Department of Condensed Matter Physics and Materials Science, Tata Institute of Fundamental Research, Homi Bhabha Road, Colaba, Mumbai 400 005, India.}
\ead{pkd@tifr.res.in}

\begin{abstract}
The equiatomic PrGe single crystal was grown by Czochralski pulling method. The grown single crystal was found to have CrB-type orthorhombic crystal structure with the space group \textit{Cmcm}~(\#63). Transport and magnetization data reveal large anisotropy in the electrical resistivity, magnetic susceptibility and magnetization. PrGe was found to exhibit two consecutive magnetic orderings at 44~K and 41.5~K, respectively. The magnetic susceptibility measurement along the three principal directions, in low applied fields, revealed a cusp like behaviour at 44~K while at 41.5~K a ferromagnetic like increase was observed. The hysteritic behaviour in the magnetization measurement at 1.8~K confirmed the ferromagnetic nature of PrGe at low temperatures. The heat capacity data clearly revealed the bulk nature of two magnetic transitions by the presence of two sharp peaks attaining values exceeding 40~J/K$\cdot$mol at the respective temperatures. The absence of Schottky contribution in the magnetic part of heat capacity indicates a quasi-ninefold degenerate $J~=~4$ magnetic ground state in this system. The low temperature data of electrical resistivity and the magnetic part of heat capacity show an existence of gap in the spin-wave spectrum.
\end{abstract}
\submitto{\JPCM}
\maketitle

\section {Introduction}
The equiatomic RX series of compounds with R as rare-earth and X as either Si, Ge, Ni or Pt crystallizing in the orthorhombic crystal strcuture, exhibit not only interesting structural properties but also show interesting magnetic properties~\cite{Nguyen, Fillion, Gomez}. One of the earliest structural studies on this RX (R = rare earth, X = Si, Ge, Ni, Pt) was performed by Hohnke and Parth\'{e}~\cite{Hohnke} about five decades ago. They found that RSi (R = La - Er) crystallize in the FeB-type structure while the higher rare earths from Dy to Er in this series exhibit polymorphic behaviour by crystallizing in CrB-type orthorhombic structure as well. On the other hand, most of the RGe compounds (R = Nd - Er) crystallize in the CrB-type crystal structure. The RGe compounds of lighter rare earths namely La and Ce crystallize in the FeB-type structure while PrGe exhibits polymorphism by crystallizing in the FeB and CrB-type orthorhombic crystal structure depending on the preparation conditions. Buschow and Fast~\cite{Buschow} investigated the magnetic properties on polycrystalline samples of RGe (R = La - Er) and reported that PrGe and NdGe order ferromagnetically while CeGe, SmGe, GdGe, TbGe, DyGe, HoGe and ErGe exhibit antiferromagnetic ordering on cooling. In EuGe, the Eu was found to be in divalent nature with an antiferromagnetic ordering at 20~K~\cite{Bushow2}.  From neutron diffraction study, Schobinger and Buschow~\cite{Schobinger} have confirmed the ferromagnetic ordering in NdGe with a Curie temperature of 28~K.   Furthermore, they stabilized  PrGe in its two polymorphic forms and reported a $T_{\rm C}$ of 36~K for FeB-type and 39~K for CrB-type orthorhombic structure with the magnetic moment pointing along the $c$-axis. Lambert-Andron \textit{et al}~\cite{Lambert}, have investigated the off-stoichiometric PrGe$_{1.66}$ compound, which was found to exist in two different crystallographic sturctures possessing the ordered and disordered variants of the ThSi$_2$ type crystal structure.  They reported that the both the ordered and disordered phases are ferromagnetic with ordering temperatures at 19 and 14~K respectively.  Recently, we investigated the anisotropic magnetic properties of CeGe single crystal in detail, which orders antiferromagntically with a N\'{e}el temperature of 10.5~K~\cite{Pranab}. The magnetic easy axis was found to be along the [010] direction. The electrical resistivity of CeGe showed an upturn at the N\'{e}el temperature indicating the superzone gap formation. In continuation to our studies on RGe compounds, and as there are no further detailed studies on the PrGe compound, we have grown the single crystal of PrGe and report the anisotorpic magnetic properties here.

\section{Experiment}

The binary phase diagram of PrGe revealed that PrGe melts congruently~\cite{Gokhale} at 1400~$^{\circ}$C, hence we adopted Czochralski crystal pulling method to grow the single crystal in a tetra-arc furnace. The starting materials were high purity metals of Pr~(99.95\%) and Ge~(99.999\%). A total of 10~g of the stoichiometric PrGe was repeatedly melted several times to make a homogeneous mixture. A thin polycrystalline seed rod was cut out of this polycrystalline material and immersed into the melt and pulled at a speed of 10~mm/h in a pure and dry argon atmosphere.  The crystal was pulled for about 6 to 7 hours to obtain 70~mm long crystal with a diameter roughly about 3 to 4 mm. The phase purity of the single crystal was investigated by means of powder x-ray diffraction using PANalytical machine. The single crystal was oriented along the principal crystallographic directions by means of Laue back reflection using a Huber Laue diffractometer and cut to the desired dimensions using a spark erosion cutting machine. For the non-magnetic reference compound, a polycrystalline sample of YGe was prepared by melting stoichiometric amounts of high purity Y and Ge in an arc-melting furnace.  The polycrystalline sample was then annealed at 1100~$^\circ$C for one week in a resistive heating furnace.  The magnetic measurements were performed using a quantum interference device (SQUID) magnetometer and vibrating sample magnetometer (VSM). The electrical resistivity measurement was performed in a home made resistivity set up in the temperature range from 1.8 to 300~K. The heat capacity measurement was performed using a physical property measurement system (PPMS).

\section{Results}

\subsection{X-ray diffraction}

Owing to the fact that PrGe exists in two polymorphic forms~\cite{Schobinger} namely the CrB and FeB-type orthorhombic crystal structure with the space groups \textit{Cmcm}~(\#63) and \textit{Pnma}~(\#62) respectively, as a first step we wanted to confirm the phase purity and the crystal structure of our single crystal. A small portion of the grown crystal was subjected to powder x-ray diffraction at 300~K using
\begin{figure}
\begin{center}
\includegraphics[width=0.8\textwidth]{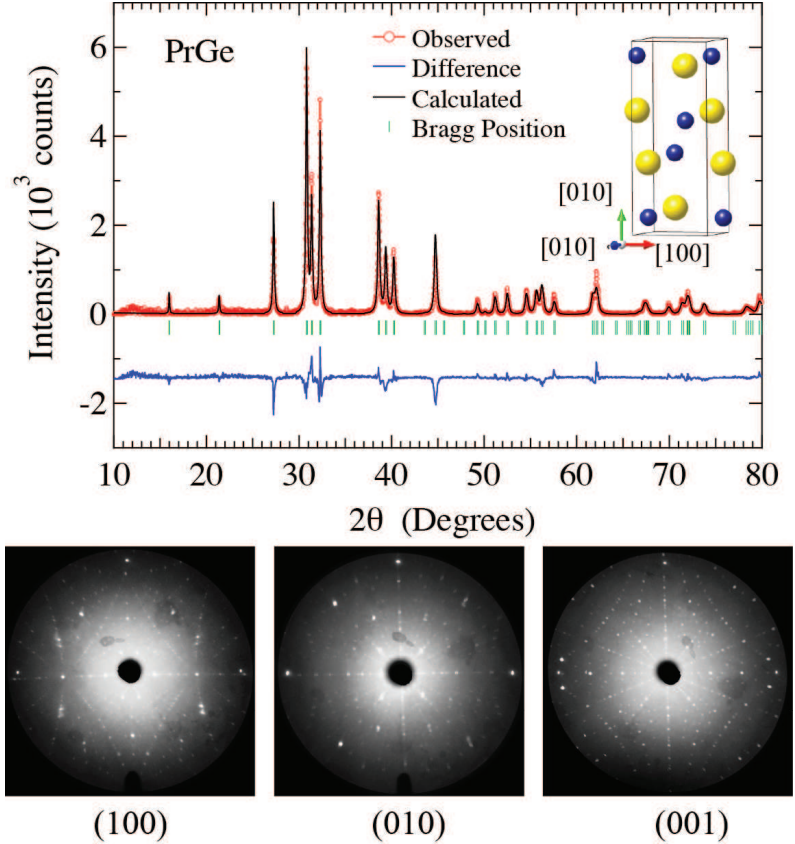}
\caption{\label{fig1} Powder x-ray diffraction pattern of PrGe together with the Rietveld refinement pattern. The inset shows the crystal structure of PrGe.  The Laue diffraction pattern corresponding to (100), (010) and (001) are also shown.}
\end{center}
\end{figure}
a monochromatic Cu-$K_{\rm \alpha}$ radiation with the wavelenth 1.5406~\AA. The x-ray pattern reveals that the grown crystal possesses the CrB-type structure as the major phase (Fig.~\ref{fig1}). A Rietveld refinement using the Fullprof program~\cite{Fullprof} confirmed the orthorhombic crystal structure of CrB-type with the space group \textit{Cmcm}. The lattice constants were estimated to be $a=4.4808 (\pm~0.5066~\times 10^{-3})$, $b=11.087 (\pm~0.1101~\times 10^{-2})$ and $c=4.050 (\pm~0.3855~\times 10^{-3})$~\AA~ with the reliability parameters $R_{\rm B}$ = 12.4~\% and $R_{\rm F}$ = 6.50~\%. The lattice constants are in close agreement with those reported in Ref.~\cite{Schobinger}. Both the Pr and Ge atoms occupy the 4c Wyckoff's position, with $y$~=~0.3608(3) and 0.0785(5), respectively. The stoichiometry of the grown crystal was further confirmed by means of energy dispersive analysis by x-ray measurement. The back reflection Laue pattern of the grown crystal was characterized by well defined Laue diffraction spots ascertaining the high quality of the crystal, and it was also employed to identify the three principal crystallographic directions as shown in Fig.~\ref{fig1}. The crystal was then cut along the principal directions by means of the spark erosion cutting machine for the anisotropic physical property measurements.

\subsection{Magnetic susceptibility and magnetization}
\begin{figure}
\begin{center}
\includegraphics[width=0.8\textwidth]{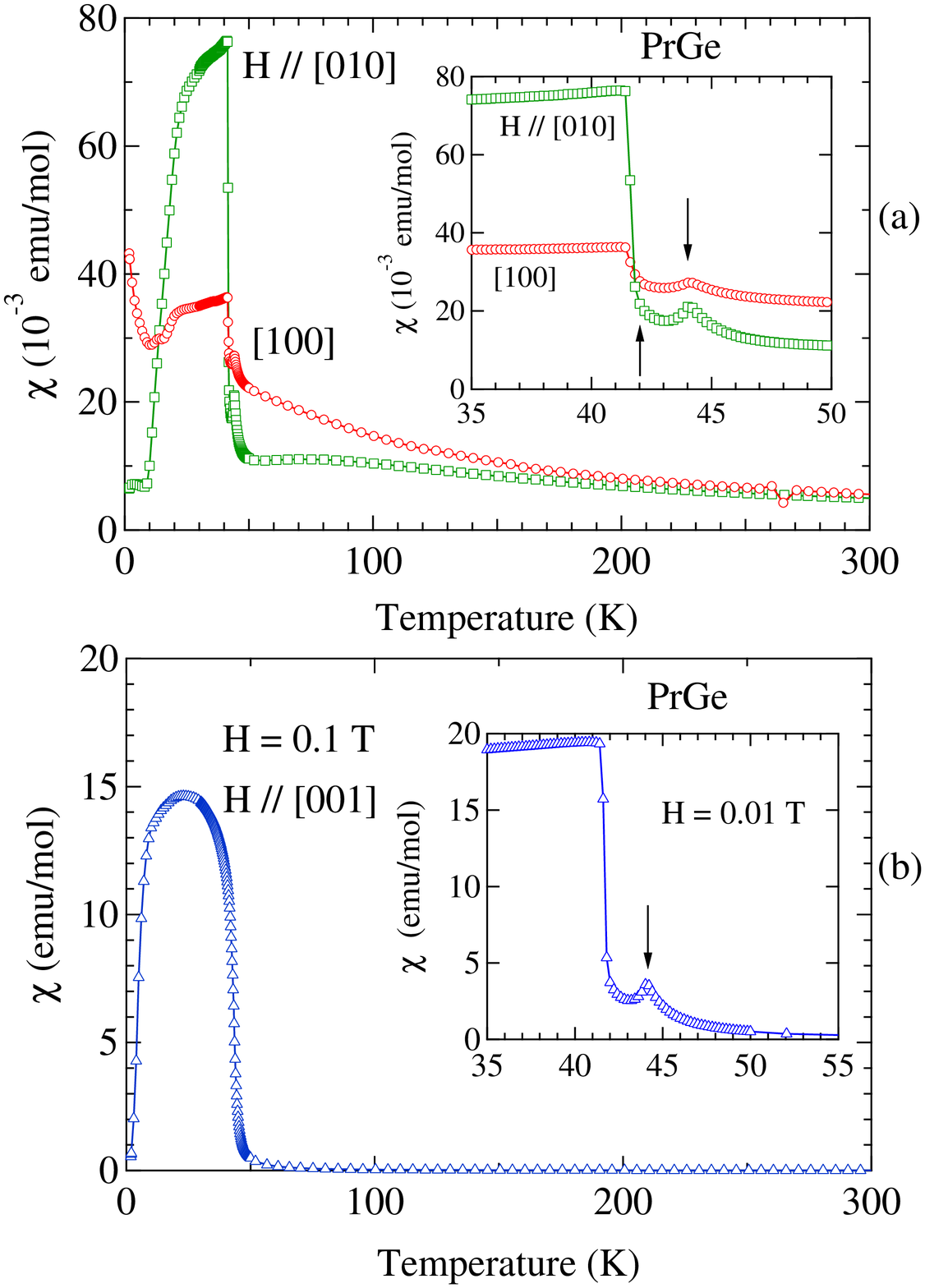}
\caption{\label{fig2} (a) Temperature dependence of magnetic susceptibility of PrGe along [100] and [010] directions, measured in an applied field of 0.1~T. (b) Susceptibility of PrGe along [001] direction measured in an applied field of 0.1~T.  The inset in both the figure shows the expanded scale susceptibility near the ordering temperature. The arrows indicate the magnetic ordering. The inset in (b) shows data measured in a field of 0.01~T.}
\end{center}
\end{figure}

Figure~\ref{fig2}(a) shows the temperature dependence of magnetic susceptibility of PrGe measured in an applied magnetic field 0.1~T in the temperature range from 1.8 to 300~K for H~$\parallel$~[100] and [010] directions. There is a large anisotropy in the 
magnetic susceptibility along the two directions. Signatures of two magnetic orderings at 44~K and at 41.5~K is observed along both [100] and [010] directions.  From the inset of Fig.~\ref{fig2}(a) it is obvious that the high temperature ordering at 44~K indicates a cusp like behaviour usually observed in the case of antiferromagnetic ordering. At 41.5~K the magnetic susceptibility shows an upturn signaling a ferromagnetic ordering. For $H~\parallel$~[100] direction, below 10~K, the susceptibility increases this may be attributed to the canting of the ordered Pr moments along this hard axis and/or due to crystallographic defects.  The magnetic susceptibility for H~$\parallel$~[001] direction is shown in Fig.~\ref{fig2}(b). Unlike the other two directions, the susceptibility in a field of 0.1~T apparently exhibits only one ordering at 44~K which is ferromagnetic like. However when the data are recorded in a lower field of 0.01~T (inset of Fig.~\ref{fig2}(b)) the antiferromagnetic ordering at 44~K is clearly observed in the [001] direction as well, followed by a ferromagnetic like upturn at lower temperatures. The in field, isothermal magnetization data to be discussed later further substantiate this claim.
\begin{figure}
\begin{center}
\includegraphics[width=0.8\textwidth]{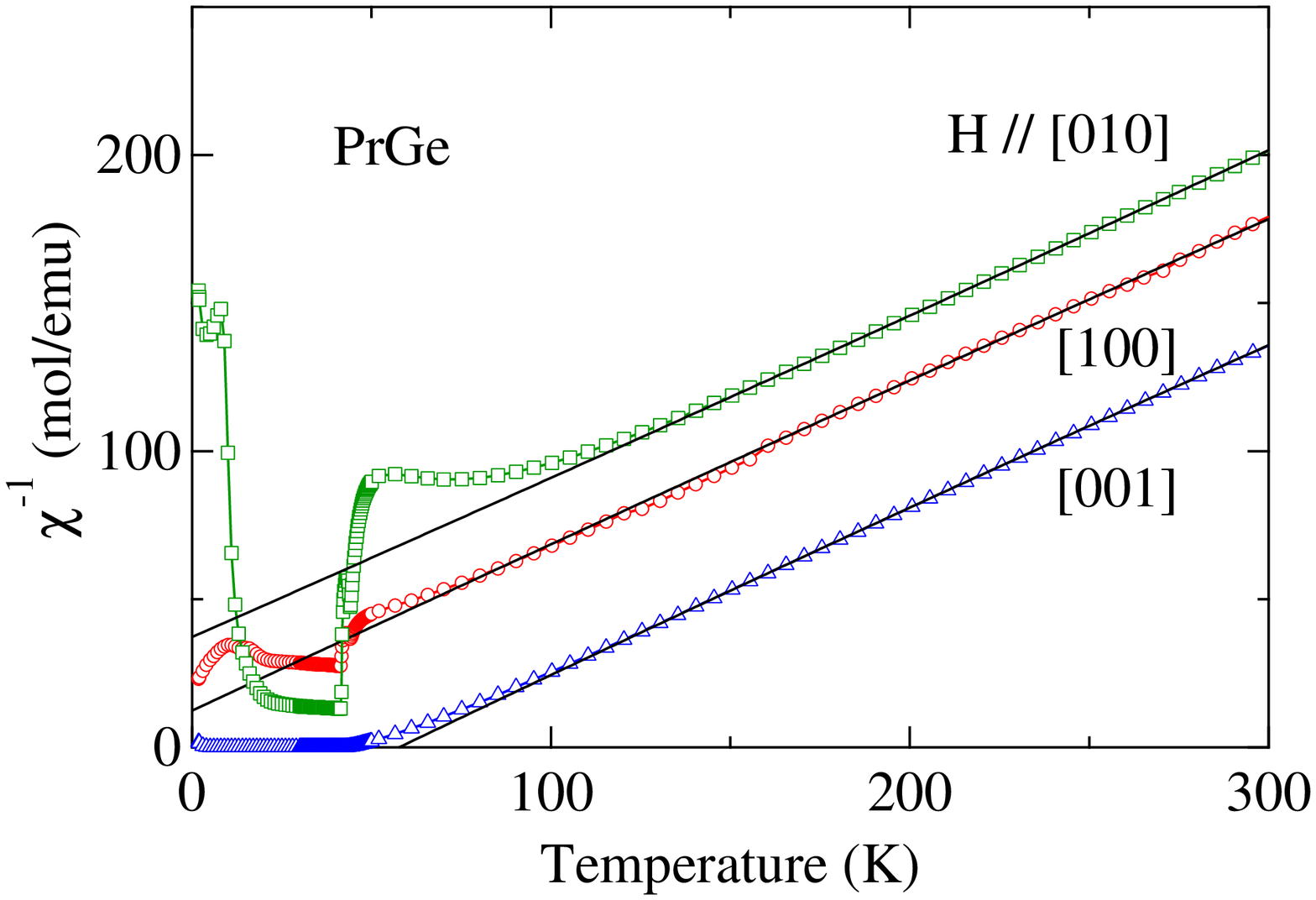}
\caption{\label{fig3} Inverse magnetic susceptibility of PrGe along the three principal crystallographic directions, the solid lines are fits to modified Curie-Weiss law.}
\end{center}
\end{figure}
The inverse magnetic susceptibility of PrGe is shown in Fig.~\ref{fig3}.  The anisotropy in the magnetic susceptibility along the three principal directions is clearly evident. The solid lines are fits to the modified Curie-Weiss law $\chi = \chi_0 + C/(T-\theta_{\rm p})$. It is observed that the Curie-Weiss law is obeyed from 300~K to immediately above the magnetic ordering and no deviation due to crystal electric field effect is observed. From the fitting, effective magnetic moment $\mu_{\rm eff}$, the paramagnetic Curie temperature $\theta_{\rm p}$ and the temperature independent magnetic susceptibility $\chi_0$ were found to be 3.78~$\mu_{\rm B}$, $-24$~K and 6.794~$\times$~10$^{-5}$~emu/mol; 3.90~$\mu_{\rm B}$, $-70$~K and 1.8166~$\times$~10$^{-4}$~emu/mol and 3.71~$\mu_{\rm B}$, 58~K and 2.5831~$\times$~10$^{-4}$~emu/mol, respectively for $H~\parallel~$[100], [010] and [001] directions. The experimental value of the effective moment $\mu_{\rm eff}$ is close to the free ion value of Pr$^{3+}$, 3.58~$\mu_{\rm B}$. The Weiss temperature  $\theta_{\rm p}$ is positive for $H~\parallel~$[001], as expected for a ferromagnetic ordering compound. On the other hand, $\theta_{\rm p}$ is negative for $H~\parallel$~[100] and [010] directions due to the antiferromagnetic ordering at 44~K.
\begin{figure}
\begin{center}
\includegraphics[width=0.8\textwidth]{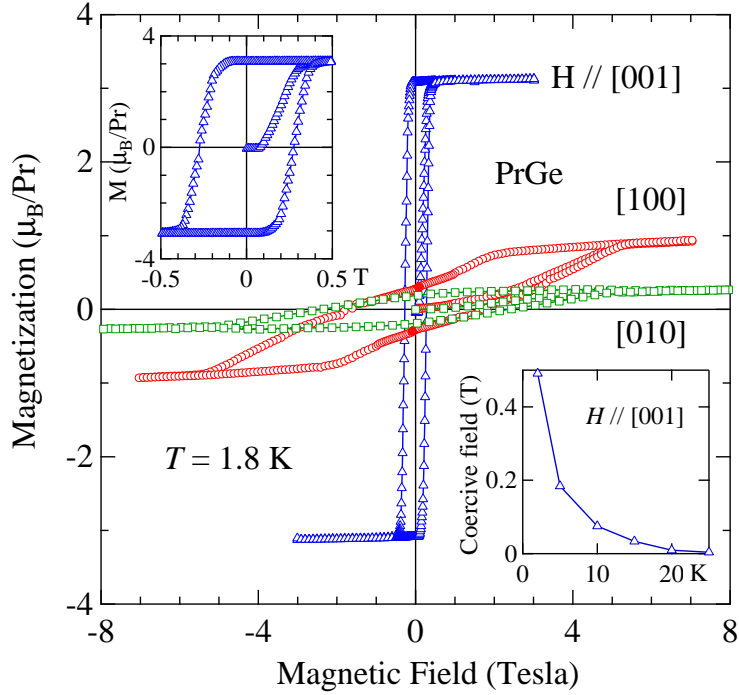}
\caption{\label{fig4}Isothermal magnetization of PrGe measured at temperature 1.8~K for fields along the three principal crystallographic directions. The upper  inset shows the low field part of the hysteresis loop for $H~\parallel$~[001] direction.  The lower  inset shows the temperature dependence of the coercive field for $H~\parallel$~[001] direction.}
\end{center}
\end{figure}
The isothermal magnetization of PrGe measured at $T=1.8$~K along the three principal crystallographic directions is shown in Fig.~\ref{fig4}. Hysteretic behaviour is observed along all the three directions confirming the ferromagnetic ground state in PrGe, thus corroborating the previous neutron diffraction results on a polycrystalline sample of PrGe~\cite{Schobinger}. For $H~\parallel$~[001] the magnetization increases more rapidly with field than along the other two directions thus indicating [001] direction as the easy axis of magnetization. At 1.8~K the magnetization saturates to 3.12~$\mu_{\rm B}$/Pr, very close to the saturation moment of free Pr$^{3+}$ ion ($g_J J \mu_{\rm B}$ = 3.20~$\mu_{\rm B}$). The saturation value of the magnetic moment along [100] direction and [010] directions are much smaller indicating the hard axes of magnetization. The width of the hysteresis loop decreases as the temperature increases and above 42~K, where the sample enters into the antiferromagnetic state, there is no signature of hysteresis.  This is shown in the lower inset of Fig.~\ref{fig4} as a plot of coercive field versus temperature for $H~\parallel$~[001] direction.   For $H~\parallel$~[001] direction the slope of the virgin curve is small which indicates the presence of narrow domain walls~\cite{Broek}. The impurities of atomic dimensions pin the domain wall as it is evident from the flat virgin curve upto a field of 0.08~T (see inset of Fig.~\ref{fig4}), at which point the magnetization increases, which means the external field is able to detach the pinned domain wall at this field of 0.08~T, the complete detaching of the pinned domain wall happens at 0.4~T and at higher fields the domain walls are completely removed. On reversing the field the reversed domain walls nucleate and only for the negative fields of about 0.4~T the domain walls are removed.

\subsection{Electrical Resistivity}

\begin{figure}
\begin{center}
\includegraphics[width=0.8\textwidth]{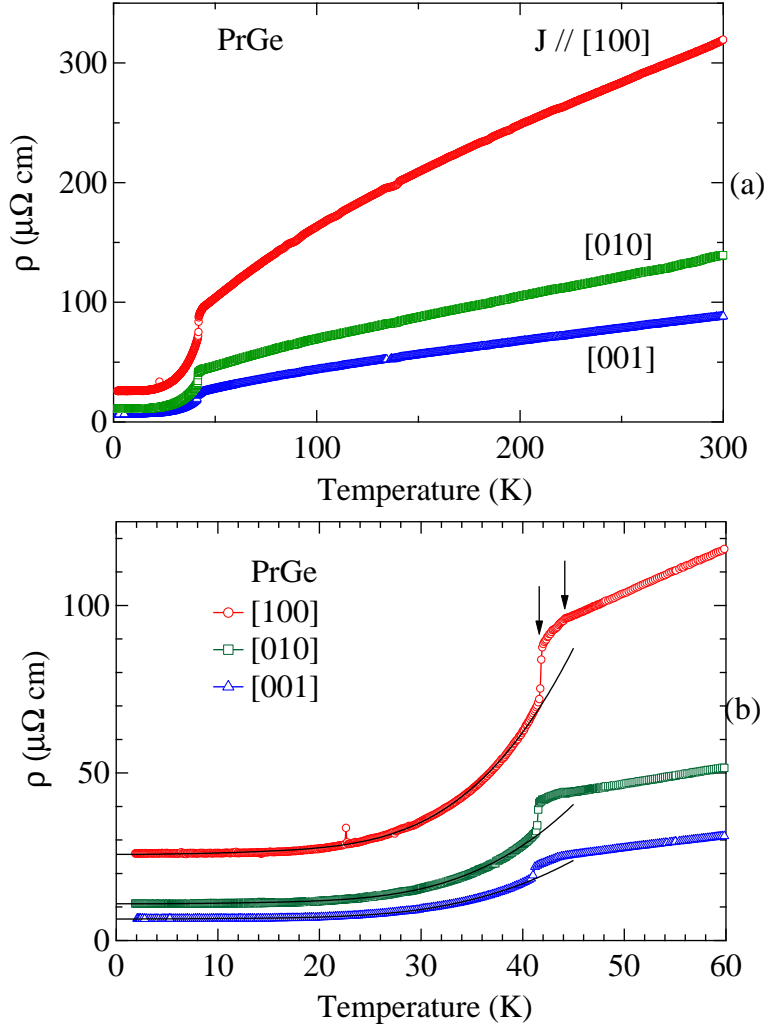}
\caption{\label{fig5}(a) Temperature dependence of electrical resistivity of PrGe for current parallel to the three principal crystallographic direction. (b) The low temperature part of the resistivity of PrGe. The solid lines are the fit to the spin-wave gap model given in Eq.~\ref{eqn1}.}
\end{center}
\end{figure}

Figure~\ref{fig5}(a) shows the temperature dependence of electrical resistivity of PrGe single crystal measured for current parallel to [100], [010] and [001] directions. The electrical resistivity decreases with decreasing temperature along the three directions and the magnetic ordering is clearly visible by the sharp drop due to the reduction in the spin disorder scattering. There is a large anisotropy in the electrical resistivity with room temperature values of $\rho_{\rm [100]}(300~{\rm K})=319 \mu \Omega \cdot$cm, $\rho_{\rm [010]}(300~{\rm K})=139 \mu \Omega \cdot$cm and $\rho_{\rm [001]}(300~{\rm K})=88 \mu \Omega \cdot$cm  
for PrGe. Figure~\ref{fig5}(b) shows the low temperature part of the electrical resistivity. The high temperature antiferromagnetic ordering $T_{\rm N}$ is observed at 44~K by a very small change of slope. The ferromagnetic ordering $T_{\rm C}$ at 41.4~K shows a sharp drop in the electrical resistivity along all the three directions. At $T_{\rm N}$ there is a decrease in the resistivity.  However, just immediately below this temperature, the Pr moments again re-orient themselves to the ferromagnetic ordering because of this, the resistivity drop is not so prominent between $T_{\rm N}$ and $T_{\rm C}$.  Once the ferromagnetic ordering sets-in the electrical resistivity drops more rapidly due to the reduction in the spin-disorder scattering.  This type of sharp drop in the electrical resistivity is attributed to the first order transition. The electrical resistivity of PrGe in the ferromagntically ordered state can be fitted to the expression
\begin{equation}
\label{eqn1}
\rho(T) = \rho_0 + aT^2 + D T \Delta \left(1 + \frac{2T}{\Delta}\right)\rm{exp}\left(-\frac{\Delta}{T}\right),
\end{equation}
which involves apart from the usual Fermi liquid term, the contribution from the energy gap in the magnon dispersion relation~\cite{Andersen}. In Eq.~\ref{eqn1} $\rho_{\rm 0}$ is the residual resistivity, $a$ is the coefficient responsible for electron-electron scattering, D is the coefficient of electron-magnon scattering and $\Delta$ is the magnitude of the spin-wave gap. The low temperature part of the electrical resistivity is very well explained by the above expression up to 40~K as seen by the solid lines in Fig.~\ref{fig5}(b). The parameters obtained from the fit  are given in the Table~\ref{Table1}.

\begin{table}
\begin{center}
\begin{tabular}{|c|c|c|c|c|}
\hline 
 & $\rho_{0}$ & a & D & $\Delta$
\tabularnewline
 & ($\mu \Omega$~cm) & ($\mu \Omega$~cm/K$^2$)  & ($\mu \Omega$~cm/K$^2$) & (K)\tabularnewline
\hline 
$\rho_{[100]}$ & 26  & 0.004 & 0.14  & 150 \tabularnewline
\hline 
$\rho{[010]}$ & 10 & 0.002 & 0.07 & 154 \tabularnewline
\hline 
$\rho{[001]}$ & 6.4 & 0.001 & 0.03 & 142\tabularnewline
\hline
\end{tabular}

\caption{\label{Table1} Fitting paramters of the resistivity data along the three principal directions described in Eqn.~\ref{eqn1}.}
\end{center}
\end{table}

The obtained values of the spin wave gap is nearly equal for all the three directions even though there is a large anisotropy in the electrical resistivity. The residual resistivity ratio (RRR) amounts to 12.22, 12.52 and 13.54 for $J~\parallel$~[100], [010] and [001] directions respectively thus indicating good quality of the single crystal.

\subsection{Heat Capacity}
\begin{figure}
\begin{center}
\includegraphics[width=0.8\textwidth]{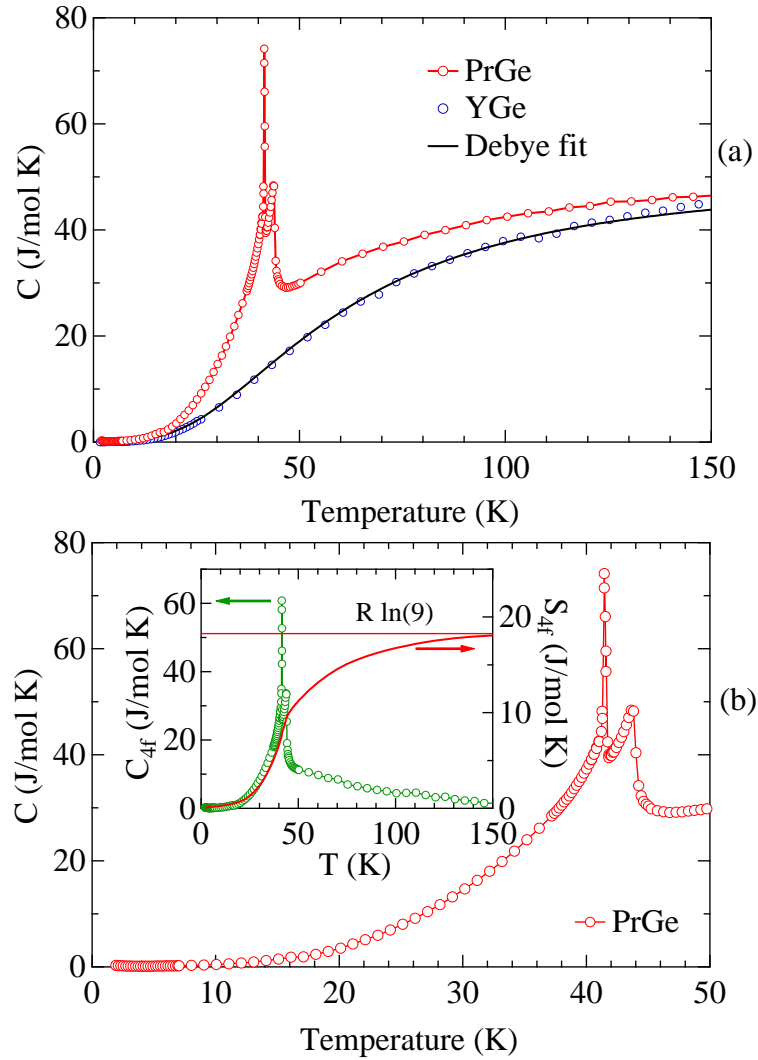}
\caption{\label{fig6}(a) Temperature dependence of the specific heat capacity in PrGe and YGe. The solid line through the data points of YGe is fit to the Debye model of heat capacity. (b) The main panel shows low temperature part of the specific heat of PrGe and the inset shows the magnetic part of heat capacity and the calculated entropy.}
\end{center}
\end{figure}

The temperature dependence of the specific heat of PrGe single crystal and its non-magnetic reference compound YGe measured in the range of 1.8 to 150~K is shown in the main panel of Fig.~\ref{fig6}(a).  Since the crystal structure of LaGe is FeB-type which is different from the CrB-type structure of PrGe, we have prepared a polycrystalline sample of YGe which possesses the same crystal structure as that of PrGe and used it as the non-magnetic reference compound after taking into account the mass renormalization as mentioned in Ref.~\cite{Bouvier}. 
The heat capacity of YGe can be very well fitted to the Debye model as shown by the solid line through the data points of YGe and $\Theta_{\rm D}$ was estimated to be 245~K.  Furthermore, an estimation of $\gamma$ was done from the low temperature heat capacity data of YGe and it was found to be 3.4~mJ/K$^{2}\cdot$mol.  The two magnetic orderings are observed exactly at the same temperatures as corresponding to their value in the  magnetic susceptibility and the resistivity data thus confirming the two bulk magnetic orderings in PrGe. Figure~\ref{fig6}(b) shows the low temperature part of  the heat capacity of PrGe, which clearly depicts both the magnetic orderings. The jump in the heat capacity at the first magnetic ordering at 44~K amounts to 19.29~J/K mol while the at the second magnetic ordering namely where the system undergoes the ferromagnetic ordering, a very sharp peak with a jump of about 45.06~J/K mol is observed. This sharp jump at 41.5~K together with the sharp resistivity drop, as mentioned earlier, clearly indicates that this is a first order transition. Further confirmation for the first order nature of the ferromagnetic transition at 41.5~K comes from the shape of the Arrott plots, (vide infra). An estimation of the magnetic part of the heat capacity of PrGe was obtained by subtracting the heat capacity of YGe which is assumed as the lattice part of the heat capacity. The temperature dependence of entropy was obtained by integrating the $C_{\rm 4f}/T$ data as shown in the inset of Fig.~\ref{fig6}(b).  
\begin{figure}
\begin{center}
\includegraphics[width=0.8\textwidth]{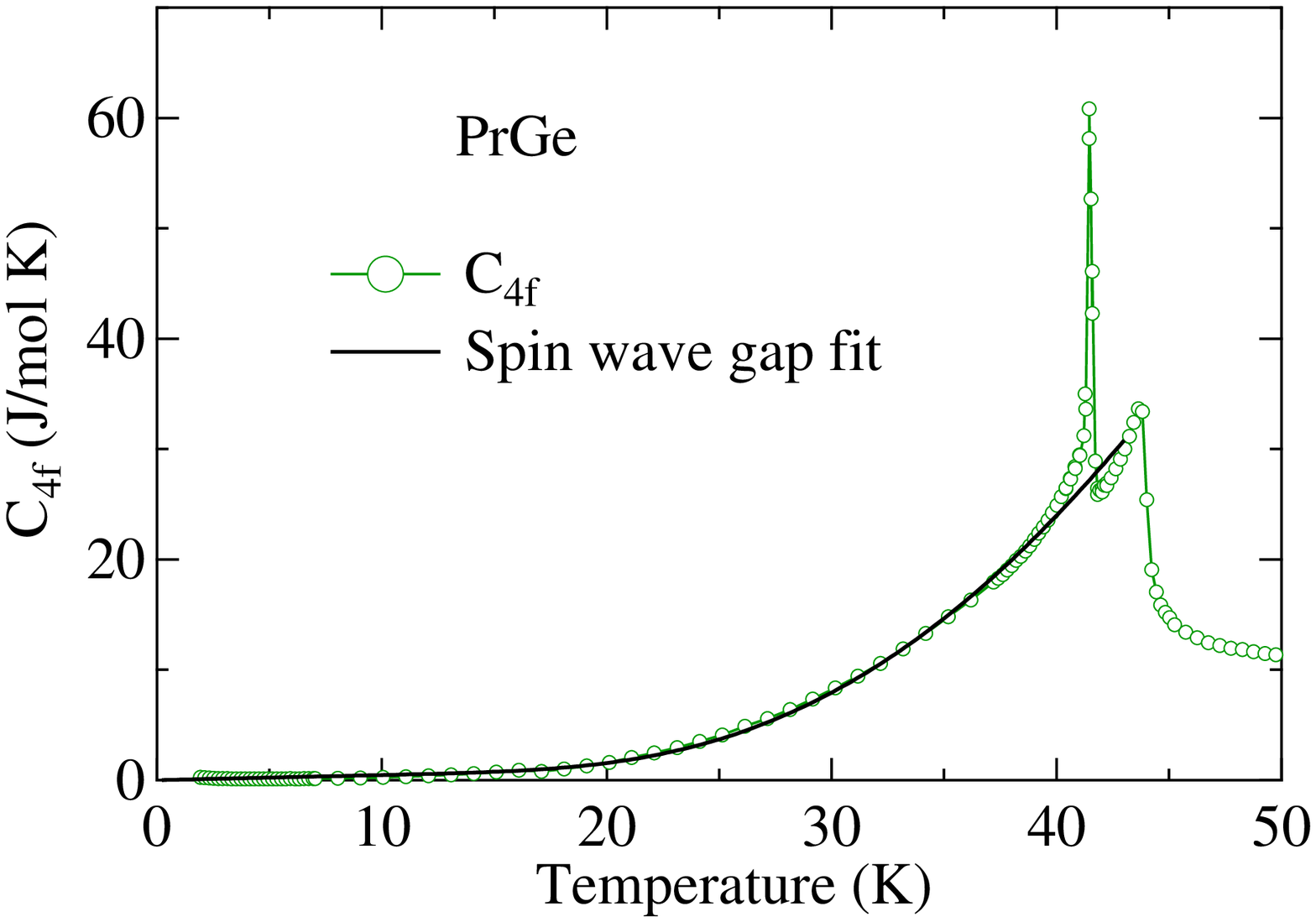}
\caption{\label{fig7}The magnetic part of the heat capacity ($C_{\rm 4f}$), the solid line is the fit to the spin-wave expression given in Eq.~\ref{eqn2}.}
\end{center}
\end{figure}
In the ferromagnetically ordered state, the magnetic contribution to the heat capacity $C_{\rm 4f}$ of PrGe can be fit to the spin-wave gap expression~\cite{Coqblin} which is given by,

\begin{equation}
\label{eqn2}
C_{\rm 4f} = \gamma T + C_{\rm SW},
\end{equation}

where
\begin{equation}
\label{eqn3}
C_{\rm SW} = \alpha \left(\frac{\Delta^2}{\sqrt{T}} + 3 \Delta \sqrt{T} + 5 \sqrt{T^{3}}\right)e^{\frac{-\Delta}{T}}.
\end{equation}

Here, $\gamma$ is the electronic term of the heat capacity and $C_{\rm SW}$ is the contribution to the ferromagnetic spin-wave excitation spectrum with an energy gap $\Delta$ and $\alpha$ is a constant. It is evident from Fig.~\ref{fig7} that the Eq.~\ref{eqn2} fits very well the $C_{\rm 4f}$ data from up to 40~K. The obtained values of the fitting parameters are $\gamma$~=~45.9~mJ/K$^2$ mol, $\alpha$~=~0.105~J/(K$^{5/2}$ mol) and $\Delta$~=~140~K. The spin-wave gap parameter $\Delta$ is in close agreement with the values obtained from the electrical resistivity data.

\section{Discussion}
The previous magnetic and neutron diffraction studies on polycrystalline samples of PrGe had revealed that PrGe orders ferromagnetically at 39~K~\cite{Buschow, Schobinger}. Furthermore, it was reported that PrGe exists in polymorhpic forms having CrB and FeB type orthorhombic crystal structure. The neutron diffraction pattern reported in Ref.~\cite{Schobinger} measured at 293~K consisted of diffraction patterns due to both CrB-type and FeB-type diffraction patterns. On the other hand, PrGe single crystal studied here did not show any polymorphism. It is evident from Fig.~\ref{fig1}, the powder diffraction pattern of PrGe single crystal measured at 300~K does not show any reflection due to the two different polymorphic structure and it matched well with the CrB-type orthorhombic crystal structure with the space group \textit{Cmcm}. Unlike the previous reports on polycrystalline samples, the transport and magnetic properties measured on PrGe single crystal in the present work reveal two consecutive magnetic orderings at 44~K and 41.5~K, respectively where the high temperature ordering is antiferromagnetic in nature while the one at 41.5~K is ferromagnetic. Hysteretic behaviour is observed along all the three crystallographic directions in the M vs. H measurements measured at 1.8~K thus confirming the ferromagnetic ground state of PrGe. In order to further confirm the
\begin{figure}
\begin{center}
\includegraphics[width=0.8\textwidth]{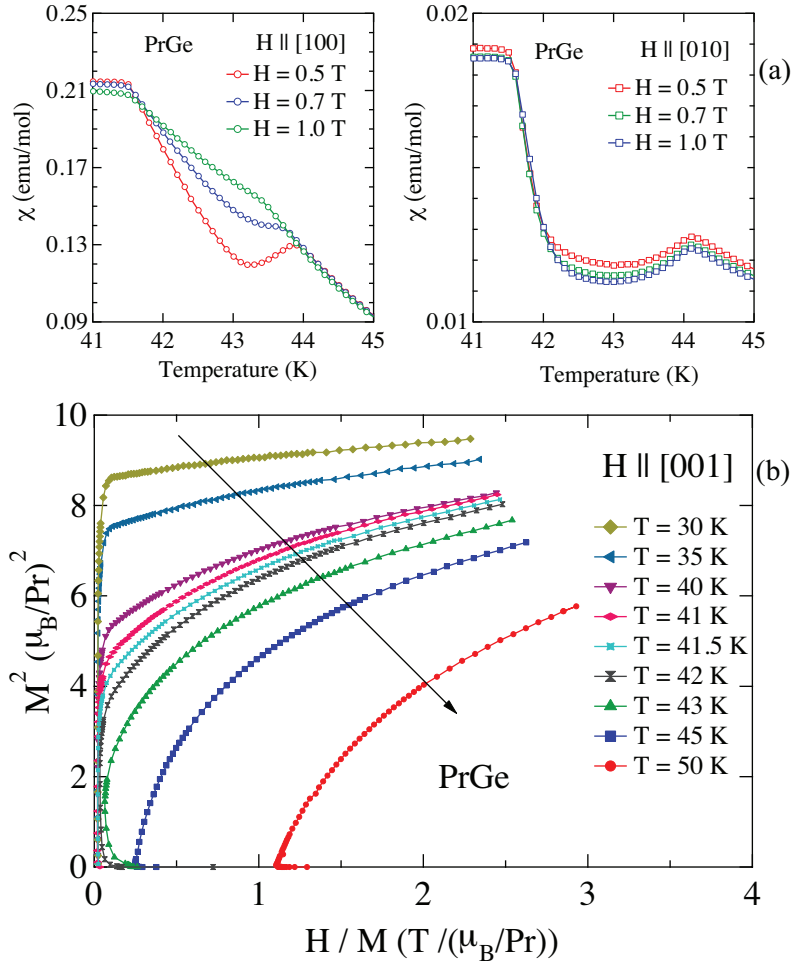}
\caption{\label{fig8}(a)Magnetic field dependence of the magnetic susceptibility of PrGe for $H~\parallel~[100]$ and [010] directions, (b)  Arrot plots of PrGe constructed in the temperature range from 30 to 50~K, the arrow indicates the progressive increase in temperature of the plots.}
\end{center}
\end{figure}
antiferromagnetic nature of the high temperature magnetic ordering at 44~K, we performed the magnetic susceptibility measurement at various applied fields viz., 0.5~T, 0.7~T and 1~T as shown in Fig.~\ref{fig8}. For $H~\parallel~$[100] direction, as the applied field increases the magnetic ordering temperature shifts to lower fields, typical for an antiferromagnet. When the field is applied parallel to [010] direction, the magnetic ordering did not show any appreciable change.  This can be attributed to the fact of antiferromagnetic hard axis of magnetization.  Since the magnetization along [001] direction increases more rapidly and saturates to 3.12~$\mu_{\rm B}$/Pr it is the easy axis of magnetization which is in conformity with the neutron diffraction result~\cite{Schobinger}. The sharp drop in the electrical resistivity and the huge jump in the heat capacity at the ferromagnetic ordering indicates a first order transition.  The Arrot plots of PrGe constructed in the temperature from 30 to 50~K is shown in Fig.~\ref{fig8}(b).  It is evident from the figure that the Arrot plots show an "S" shaped curve which is usually observed when the ferromagnetic ordering is first order~\cite{Singh}. It was first pointed out by Banerjee~\cite{banerjee} that the sign of the slope of the isotherm plots $M^2$ vs. $H/M$ is negative for a first order ferromagnetic transition. The low temperature part of the electrical resistivity and the magnetic part of the heat capacity confirm the gap in the spin-wave spectrum. In the orthorhombic site symmetry, the 9 fold degenerate Hund's rule derived ground state $^3H_4$ of Pr will be split by the crystal electric field. The orthorhombic point symmetry of Pr atoms makes this degenerate levels to split up into 9 singlets. Since PrGe orders magnetically, the low lying levels with small separation may form into an effective doublet resulting in the magnetic ordering. Just above the magnetic ordering the entropy reaches $R~ln~4$ indicating four low lying levels.  At high temperature, the magnetic entropy attains the full theoretical value of $R~ln~9~=~18.27$~J/K$\cdot$mol.  The absence of Schottky contribution  to the heat capacity, (inset of Fig.~\ref{fig6}(b)) signals the fact that the ground state may be a fully degenerate $J~=~4$ multiplet .  Similar behaviour is observed in PrSi system, in a recent report, where the Schottky contribution to the heat capacity is absent~\cite{Snyman}. To understand the complex magnetic behaviour exhibited by PrGe, a detailed neutron diffraction study on single crystal is necessary and is planned for the future.

\section{Conclusion}

Single crystals of PrGe was grown by Czochralski pulling method in a tetra-arc furnace. X-ray analysis of the sample  indicated the CrB-type structure as the only phase.  The transport and the magnetic measurements  clearly indicate that PrGe exhibit two magnetic orderings at 44~K and 41.5~K which confirmed as antiferromagnetic and ferromagnetic, respectively. A strong anisotropy along the three principal crystallographic directions was observed reflecting the orthorhombic symmetry of the crystal structure. The [001] direction was found to be the easy axis of magnetization with a complete saturation of the Pr moment at about 0.4~T. The magnetization data measured at 1.8~K revealed hysteritic behaviour along  all the three directions thus confirming the ferromagnetic nature of the configuration at this temperature. The absence of Schottky contribution to the specific heat in the magnetic part of the heat capacity indicates a possible $J$~=~4 multiplet ground state.

\end{document}